\documentclass[preprint,12pt]{elsarticle}

\usepackage{amssymb}
\usepackage{amsmath}
\usepackage{subfig}
\biboptions{sort&compress}

\journal{Optics Communications}

\newcommand{\sech}{\ensuremath{\mathrm{sech\ }}}

\begin{document}

\begin{frontmatter}

%% Title, authors and addresses

%% use the tnoteref command within \title for footnotes;
%% use the tnotetext command for the associated footnote;
%% use the fnref command within \author or \address for footnotes;
%% use the fntext command for the associated footnote;
%% use the corref command within \author for corresponding author footnotes;
%% use the cortext command for the associated footnote;
%% use the ead command for the email address,
%% and the form \ead[url] for the home page:
%%
%% \title{Title\tnoteref{label1}}
%% \tnotetext[label1]{}
%% \author{Name\corref{cor1}\fnref{label2}}
%% \ead{email address}
%% \ead[url]{home page}
%% \fntext[label2]{}
%% \cortext[cor1]{}
%% \address{Address\fnref{label3}}
%% \fntext[label3]{}

\title{Hamiltonian and Phase--Space Representation of Spatial Solitons}

%% use optional labels to link authors explicitly to addresses:
%% \author[label1,label2]{<author name>}
%% \address[label1]{<address>}
%% \address[label2]{<address>}

\author[EECS]{Hanhong Gao\corref{cor1}}
\cortext[cor1]{Corresponding author}
\ead{gaohh87@mit.edu}
\author[ME,Berkeley]{Lei Tian}
\author[ME,SMART]{George Barbastathis}

\address[EECS]{Department of Electrical Engineering and Computer Science,
 Massachusetts Institute of Technology, 77 Massachusetts Avenue, Cambridge, MA 02139, USA.}
\address[ME]{Department of Mechanical Engineering, Massachusetts Institute of Technology, 77 Massachusetts Avenue, Cambridge, MA 02139, USA.}
\address[Berkeley]{Department of Electrical Engineering and Computer Sciences, University of California Berkeley, 550 Cory Hall, Berkeley, CA 94720, USA}
\address[SMART]{Singapore--MIT Alliance for Research and Technology (SMART) Centre, 1 CREATE Way, CREATE Tower, Singapore 138602, Singapore.}

\begin{abstract}
%% Text of abstract
We use Hamiltonian ray tracing and phase--space representation to describe the propagation of a single spatial soliton and soliton collisions in a Kerr nonlinear medium. Hamiltonian ray tracing is applied using the iterative nonlinear beam propagation method, which allows taking both wave effects and Kerr nonlinearity into consideration. Energy evolution within a single spatial soliton and the exchange of energy when two solitons collide are interpreted intuitively by ray trajectories and geometrical shearing of the Wigner distribution functions.
\end{abstract}

\begin{keyword}
%% keywords here, in the form: keyword \sep keyword
Spatial soliton \sep soliton collision \sep Hamiltonian ray tracing \sep Wigner distribution function
%% MSC codes here, in the form: \MSC code \sep code
%% or \MSC[2008] code \sep code (2000 is the default)

\end{keyword}

\end{frontmatter}

%%
%% Start line numbering here if you want
%%
% \linenumbers

%% main text
\section{Introduction}

Spatial solitons, where optical beams travel without divergence or convergence in a nonlinear medium, have been theoretically presented and experimentally demonstrated in various physical systems~\cite{Chen2012,Akhmediev1997,Boyd2008,Polturak1981,Mollenauer1980,Liu2007}. Many applications have been proposed for solitons and their interactions, including optical--fiber communication systems~\cite{Blow1983}, ``gateless'' computers~\cite{Steiglitz2000}, soliton navigation~\cite{Christodoulides2003}, etc. Although theoretical methods, e.g. inverse scattering theory, exist for a few special cases~\cite{Shabat1972}, in others, it is not an easy task to predict soliton's behavior. Thus, one has to use numerical methods, such as split--step Fourier method~\cite{Agarwal2007}. In particular, energy exchange during soliton collisions has been under extensive research~\cite{Shabat1972,Gordon1983}, but much remains unknown, especially the detailed evolution of power flow during collisions. In this article, we present a novel perspective on the propagation of spatial solitons and the energy interactions during multi--soliton collision, using ray tracing and phase--space representations.

Since rays represent power flows, ray diagrams are physically intuitive and provide useful insights for the evolution of energy during a nonlinear optical phenomenon. In addition, ray tracing is easy to interpret with traditional optical terms such as ray--intercept plots, aberrations, etc. As a result, we expect ray description to be highly beneficial for understanding complex nonlinear phenomena. However, traditional ray tracing method cannot take wave effects such as diffraction and interference into consideration~\cite{Keller1962}. Furthermore, solving ray--tracing equations in Kerr nonlinear media is not straightforward because of the coupling between optical intensity and refractive index. In this article, we propose to calculate ray trajectories using the iterative nonlinear beam propagation method~\cite{Gao2010}. This method provides a rigorous way to include both wave effects and nonlinearity into the ray--tracing results. Wave effects are considered by applying the Wigner distribution function (WDF) to Hamiltonian ray tracing as the initial condition of the rays. Kerr nonlinearity, where the refractive index changes according to the local optical intensity~\cite{Swartzlander1992}, is included by an iterative process which updates the refractive index and intensity profiles at each iteration. The WDF~\cite{Walther1968,Wolf1978,Bastiaans1979} is a phase--space representation of the coherence property of an optical beam. It defines a generalized ray picture, known as the generalized radiance, which is function of position and momentum~\cite{Alonso2001}. Along each ray, the radiance is conserved~\cite{Bastiaans1979}. The optical intensity at any point of space can be calculated from the WDF through a projection along the momentum direction. The iterative nonlinear beam propagation method has been previously shown as a versatile tool for the design of nonlinear optical devices~\cite{Gao2011}. Here we show that the same method can provide useful physical insight of spatial soliton's propagation, collision and evolution with the use of ray diagrams and rigorous consideration of wave effects through the WDF.

In this article, Hamiltonian ray diagrams and phase--space representations of spatial solitons and multi--soliton collisions are studied. Energy evolution is discussed through the spatial trajectories of rays. Here, only the propagation of a single spatial soliton and the collision of two solitons are shown as examples; the same ray tracing and phase--space representation approach can be easily applied to other complex nonlinear phenomena. Furthermore, such ray representation may also be applied to the study of temporal nonlinear phenomena. For example, since spatial propagation of spatial solitons is analogous to temporal evolution of temporal solitons through a direct mapping between the space and time variables, ray tracing results presented here may be extended to temporal solitons by straightforward modifications.

\section{Spatial soliton description}

To investigate the Hamiltonian properties of a spatial soliton, we first show that given the known refractive index profile of the nonlinear medium where the soliton propagates, Hamiltonian ray trajectories yield a self--consistent result. More specifically, we show that at any given plane transverse to the optical axis, all rays have traveled for the same optical path length (OPL); moreover, the intensity distribution $I(x)$ [and thus the index profile according to the Kerr effect relation $n(x)=n_0+n_2I(x)$] maintains the same profile.

\begin{figure}[htbp]
\centering
\subfloat[]{ \includegraphics[width=12cm]{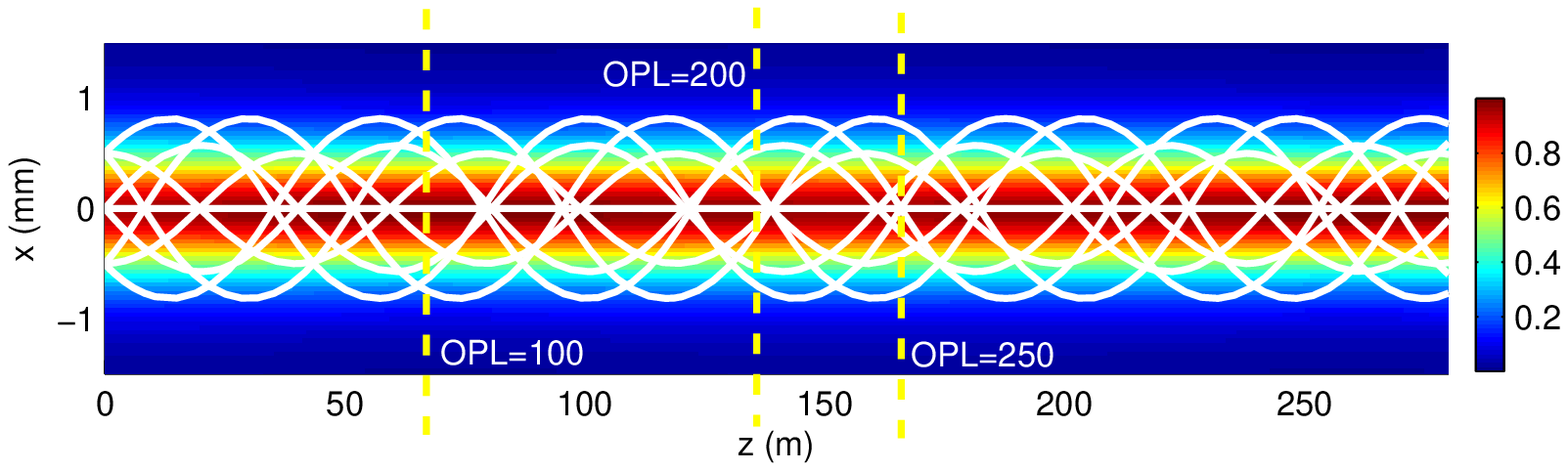} } \\
\subfloat[]{ \includegraphics[width=12cm]{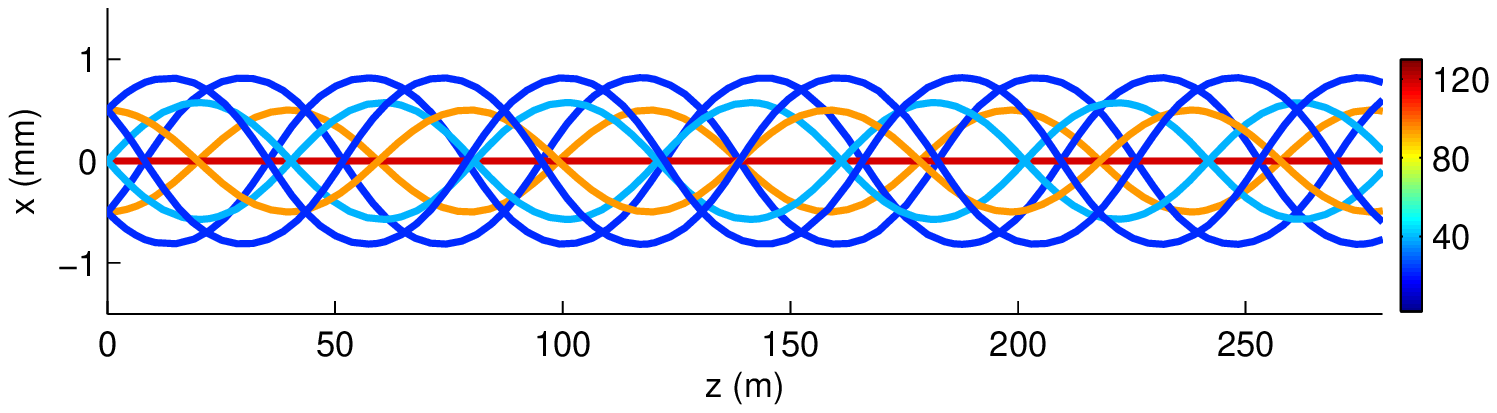} }
\caption{(a) Hamiltonian ray tracing results based on the known index profile of a spatial soliton, and (b) iterative nonlinear beam propagation method results starting from a medium of constant weak--field index. Solid lines are a subset of all $10,100$ rays used in the simulation. In (a), color shading denotes the distribution of the normalized intensity profile, proportional to the nonlinear index change $n_2I(x)$. Dashed vertical lines indicate the wavefronts with respect to different OPLs. In (b), colors of lines indicate the generalized radiances carried by these rays.}
\label{fig:SolitonHamiltonian}
\end{figure}

Hamiltonian equations describe a ray trajectory by its position $x$ and momentum $p_x$ along $x$ direction at any transverse plane $z$, for a given index distribution $n(x,z)$, and can be written as~\cite{Wolf2004}
\begin{eqnarray}
\frac{\text{d}x}{\text{d}z} = \frac{\partial h}{\partial p_x} = -\frac{p_x}{h}, \
\frac{\text{d}p_x}{\text{d}z} = -\frac{\partial h}{\partial x} = -\frac{n}{h} \frac{\partial n}{\partial x},
\label{eq:Hamiltonian}
\end{eqnarray}
\noindent where $h=-\sqrt{n^2-p_x^2}$ is the screen Hamiltonian. Note that the momentum is proportional to the direction of ray propagation by $p_x = \sin\phi/\lambda$, where $\phi$ is the angle of the propagation with respect to $z$ axis, and $\lambda$ is the wavelength. Based on the nonlinear Schr\"odinger equation, there exists an analytical solution for the optical field of a spatial soliton,
\begin{eqnarray}
A(x,z)=A_0 \ \sech(x/w_0) \ \exp{(i\theta(z))},
\label{eq:singleSoliton}
\end{eqnarray}
\noindent where $A_0$ is the peak amplitude, $w_0$ is the beam width and $\theta$ is the phase which is invariant along $x$ direction~\cite{Kivshar2003}. In a Kerr nonlinear medium, the refractive index changes proportional to the intensity distribution; thus the index profile for the spatial soliton is
\begin{eqnarray}
n(x,z)=n_0+n_2A_0^2\sech^2(x/w_0),
\label{eq:solitonIndex}
\end{eqnarray}
\noindent where the $n_0$ is usual, weak--field refractive index, and $n_2$ is the Kerr effect coefficient. Given the index distribution, the ray trajectories can be obtained by solving the pair of Hamiltonian equations in Eq.~\eqref{eq:Hamiltonian}. To obtain an input ray distribution consistent to the field description in Eq.~\eqref{eq:singleSoliton}, we compute the WDF $\mathcal{W}(x,p_x)$ of $A(x,z=0)$ to define the initial rays for Eq.~\eqref{eq:Hamiltonian}, according to the definition~\cite{Bastiaans1979}
\begin{eqnarray}
\mathcal{W}(x,p_x) = \int A(x+\frac{x'}{2})A^*(x-\frac{x'}{2})e^{-i p_x x'} \mathrm{d}x'.
\end{eqnarray}
\noindent In the simulation shown in Fig.~\ref{fig:SolitonHamiltonian}(a), we used $n_0 = 1.5$, $n_2=2\times10^{-13} \ \mathrm{(m/V)^2}$, $A_0=281 \ \mathrm{V/m}$ and $w_0=0.55 \ \mathrm{mm}$. Note that although each ray takes a distinct periodic trajectory, the rays propagate around the central region of the soliton experience higher refractive indices; the total OPLs [defined as the path integral of $n(x,z)$ along a ray trace] of all the rays at any transverse plane are the same. This result suggests that the wavefronts are always perpendicular to the optical axis which agrees with the definition of $\theta$ in Eq.~\eqref{eq:singleSoliton}.

Next we consider a dynamic process where the initial refractive index is a constant $n_0$, and show that given the initial rays satisfying the fundamental soliton solution [in Eq.~\eqref{eq:singleSoliton}] at the input plane of a Kerr medium, the solution to the Hamiltonian equations converges to the same index distribution as Eq.~\eqref{eq:solitonIndex}. We demonstrate this result using our iterative nonlinear beam propagation method. The method starts with a medium of constant weak--field refractive index $n_0$, and the definition of all the initial rays, i.e. initial position and direction, based on the WDF of the input ``sech'' profile. Each ray carries a generalized radiance, given by the value of WDF at the given position and momentum. At each iteration, we apply Hamiltonian ray tracing for each ray for the current index distribution; at the end of each iteration, the intensity at each point of space is calculated as the sum of the generalized radiances carried by all the rays passing through the point, according to the projection property of the WDF. Refractive index distribution is then updated according to the Kerr effect, whose result is used in the next iteration. As the iterations continue, all the rays converge to form a soliton. The converged ray trajectories are shown in Fig. \ref{fig:SolitonHamiltonian}(b), which match the result in Fig.~\ref{fig:SolitonHamiltonian}(a). Note that the intensity profile [i.e. the refractive index profile according to $n(x)=n_0+n_2I(x)$] is the same as Fig.~\ref{fig:SolitonHamiltonian}(a), thus we are not showing it again.

\begin{figure}[htbp]
\centering
\subfloat[]{ \includegraphics[width=6cm]{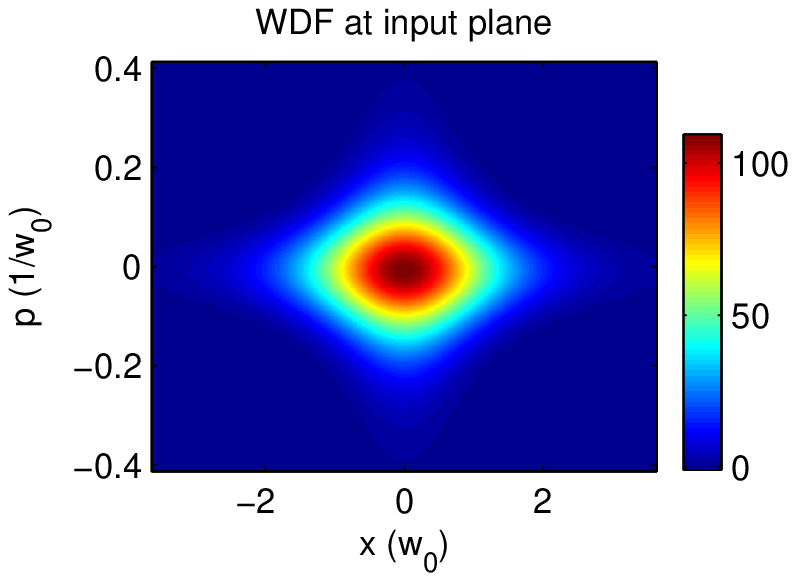} }
\subfloat[]{ \includegraphics[width=6cm]{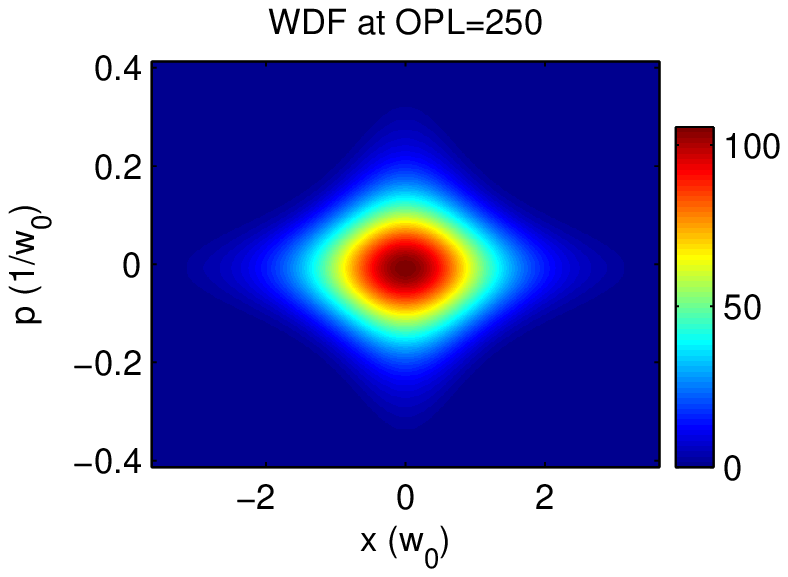} }\\
% \subfloat[]{ \includegraphics[width=5cm]{WDFDifference.eps} }
\centering
\subfloat[]{ \includegraphics[width=6cm]{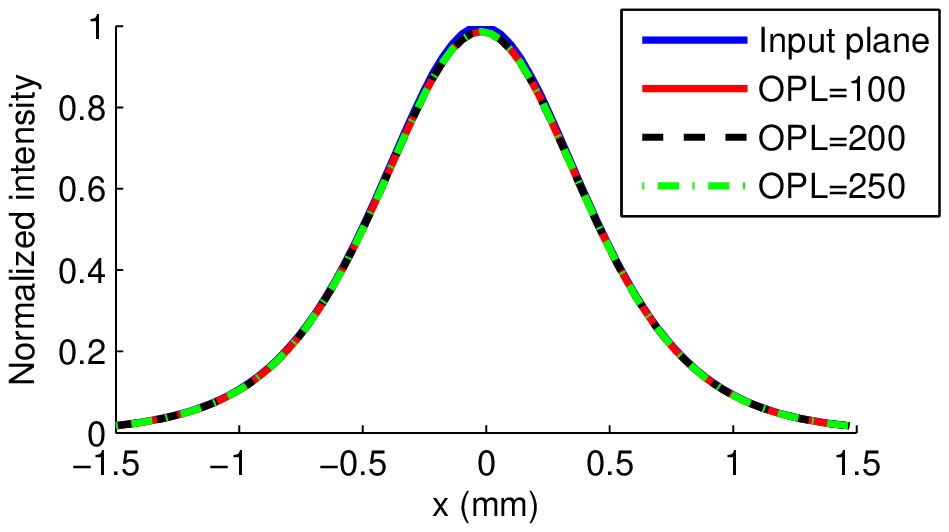} }
\subfloat[]{ \includegraphics[width=6cm]{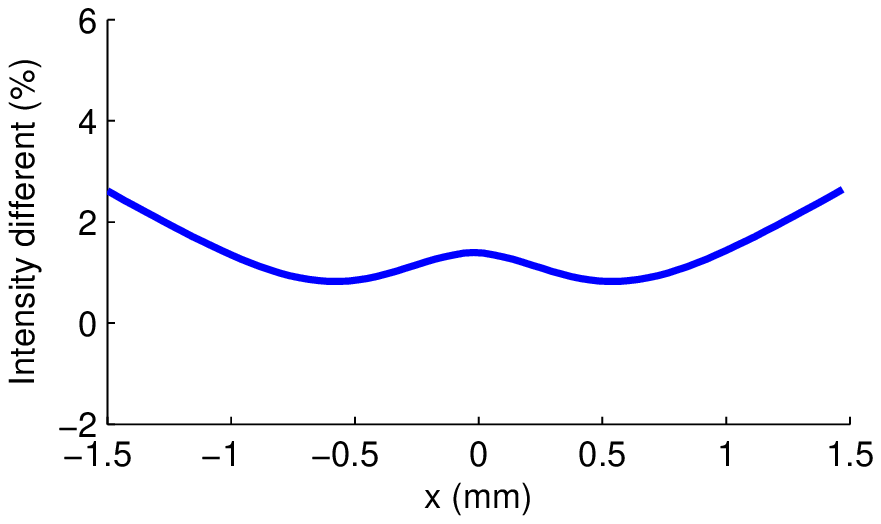} }
\caption{(a) WDF at the input plane. (b) WDF at the plane where OPL$=250$. (c) Intensity profiles at different $z$ planes. (d) The intensity difference between input plane and OPL$=250$ plane. Horizontal axis $x$ is the transversal position and $p$ is the momentum, i.e. direction of ray propagation.%(c) The difference between (a) and (b).
}
\label{fig:WDFDiagrams}
\end{figure}

According to the ray tracing results, rays with different generalized radiances and initial condition oscillate at different periods. Though most of the rays propagate in oscillatory fashion instead of straight lines parallel to the optical axis, the generalized radiances of all rays sum up to the correct intensity profile of a spatial soliton. The WDFs calculated from the rays intercepting two different $z$ planes are shown in Fig.~\ref{fig:WDFDiagrams}. As illustrated in the figure, both the WDF and the intensity distribution remains invariant as the soliton propagates, which matches the Hamiltonian ray tracing description and also the analytical results.

\begin{figure}[htbp]
\centering
\subfloat[]{ \includegraphics[width=8.67cm]{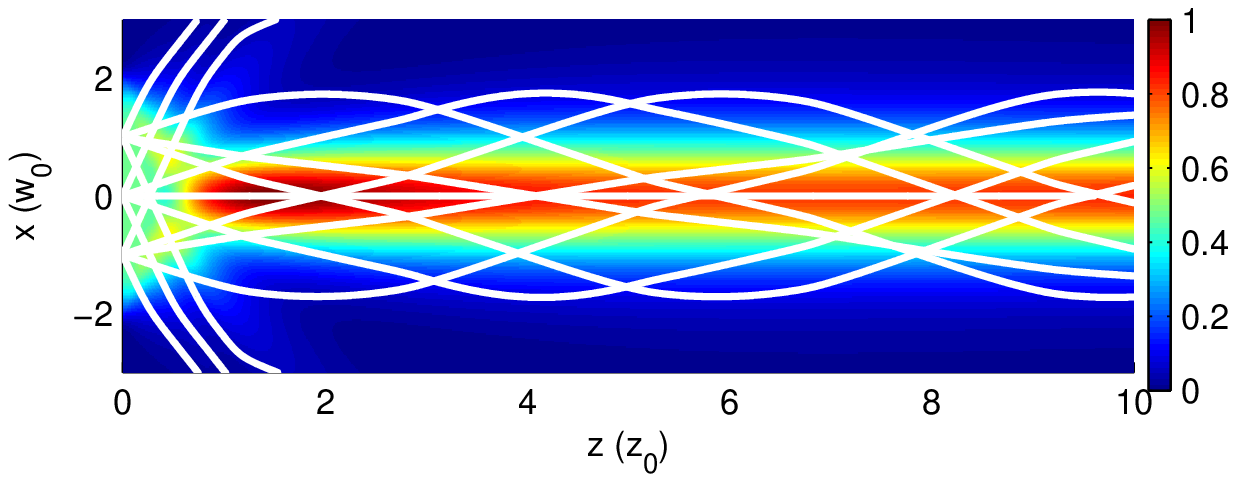} }
\subfloat[]{ \includegraphics[width=4.33cm]{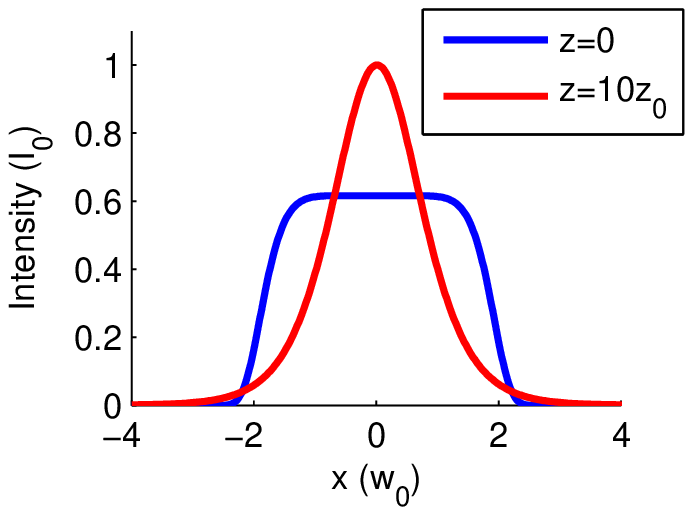} }
\caption{(a) Spatial soliton formation from super--Gaussian input calculated using our iterative method. White lines are a subset of all rays used in the simulation. (b) Transversal intensity profiles at the input and output $z$ planes.}
\label{fig:SuperGaussian}
\end{figure}

Next we consider an input deviated from the ideal fundamental soliton shape. Theoretically, it has been predicted that given an appropriate perturbation, the beam will automatically evolve into a fundamental soliton~\cite{Agarwal2007}. Here we choose super--Gaussian~\cite{Gavish2008} as a perturbed input example. A super--Gaussian with optical field $A(x) = 0.785\cdot\exp(-(x/(2w_0))^8)$ is launched from the initial plane into the Kerr nonlinear medium. The simulation starts with a medium of constant refractive index $n_0$. The converged solution from the iterative method is shown in Fig.~\ref{fig:SuperGaussian}, where it is observed that the beam adjusts its shape and width, and becomes a fundamental spatial soliton, i.e. with a ``sech'' profile. In this process, most of the energy is coupled into the soliton while some energy spreads out as leaky waves \cite{Agarwal2007}. This is shown more intuitively in terms of ray trajectories, where six sampled rays spread away from the soliton [see Fig.~\ref{fig:SuperGaussian}(a)]. The rays that are coupled into the soliton propagate in oscillatory fashion with different periods. These results agree with spatial soliton ray tracing described above. The iterative method estimates that $8.3\%$ of input energy is lost, which is close to the result calculated from the split--step Fourier method ($7.9\%$)~\cite{Agarwal2007}.

\section{Soliton collision}

\begin{figure}[htbp]
\centering
\subfloat[]{ \includegraphics[width=7cm]{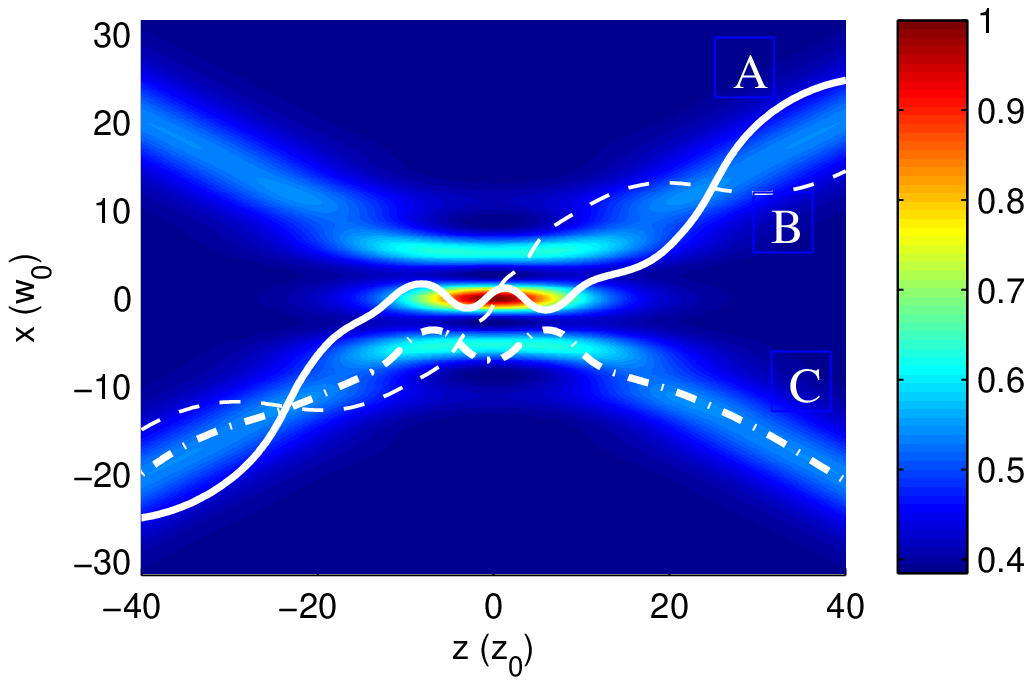} }
\subfloat[]{ \includegraphics[width=6.5cm]{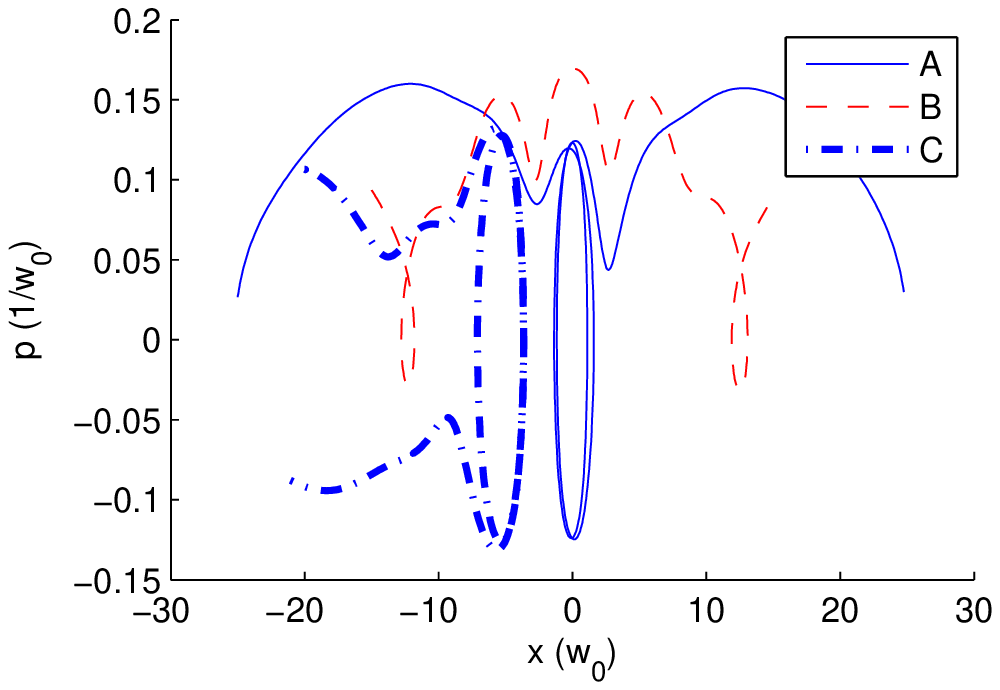} }
\caption{(a) Intensity profile (normalized) and three sampled ray trajectories of two colliding spatial solitons. (b) Evolution of the three sampled rays represented in phase--space diagram (WDF). Here $z_0 = 0.55~\mathrm{mm}$.}
\label{fig:SolitonCollision}
\end{figure}

\begin{figure}[htbp]
\centering
\includegraphics[width=12cm]{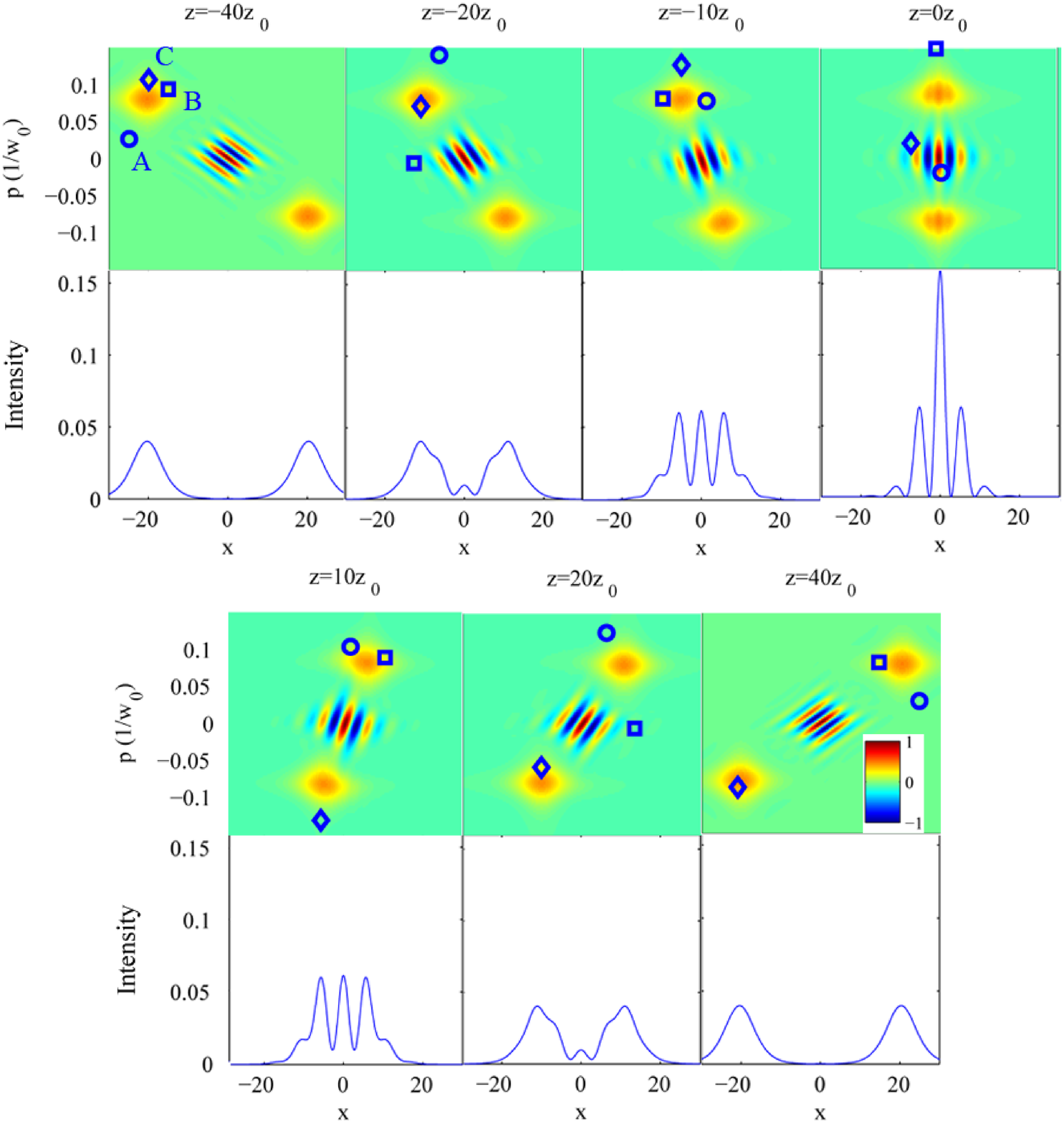}
\caption{The WDFs (top) and transversal intensity profiles (bottom) of two spatial soliton collision at different $z$ positions. Marks on the WDF correspond to the three sampled rays in Fig.~\ref{fig:SolitonCollision}, where Ray A is denoted as $\circ$, B as $_\square$, and C as $\diamond$.}
\label{fig:CollisionWDf}
\end{figure}

Next we proceed with the analysis of soliton collisions using our iterative nonlinear beam propagation method and the WDF. Collision of two spatial solitons has been solved analytically using the nonlinear Schr\"odinger equation~\cite{Akhmediev1997,Aossey1992,Stegeman1999}; however, it does not give an intuitive picture of how the energy exchanges during the collision. Here, without loss of generality, we take two solitons with the same peak amplitude and no initial phase difference as the input to a Kerr nonlinear medium. Initial rays are defined according to the WDF of the analytical optical field at the input plane expressed in Eq.~(3.78) of Ref.~\cite{Akhmediev1997}, where $a=0.25/w_0$ and $b=0.1/w_0$. Simulation results using the iterative nonlinear beam propagation method are shown in Fig.~\ref{fig:SolitonCollision}. Corresponding WDFs for selected $z$ positions are shown in Fig.~\ref{fig:CollisionWDf}, together with intensity profiles calculated from a projection of the WDF along momentum direction. It is shown that the collision process can be intuitively interpreted as shearing along the $x$ direction of the WDF, as expected from the propagation property of the WDF~\cite{Bastiaans1979}. The WDF consists of two bright ``spots'' (solitons) and an ``interference'' pattern due to the nature of coherent light~\cite{Alonso2001}. Collision occurs when the two ``spots'' line--up along the momentum direction. During the collision, the oscillating ``interference'' pattern in the middle of the WDF results in a multi--peak intensity profile; while way from the collision, the ``interference'' pattern is tilted at a large enough angle so that the oscillation adds up to zero when summing along the momentum direction, resulting in two distinct solitons. In this way, the WDF clearly explains the ``interference'' of two solitons during the collision. Fig.~\ref{fig:SolitonCollision} also plots three ray trajectories as examples of all $20,200$ rays used. These rays are also represented on the phase--space diagram to better illustrate their behavior [see Fig.~\ref{fig:SolitonCollision}(b) and the marks on Fig.~\ref{fig:CollisionWDf}]. Ray B remains in the top ``spot'', while Ray A and C oscillate between two ``spots'' during the collision; this is a clear indication of energy exchange during the collision. After the collision, Ray A remains at the original soliton but Ray C switches to the other. These results agree with the ray trajectories in Fig.~\ref{fig:SolitonCollision}(a), providing more insights into the energy interactions during the soliton collision. In addition, although energy exchange exists in the ray diagrams, the \textit{net} power transfer between the two solitons is zero, which agrees with analytical prediction~\cite{Akhmediev1993} and experimental observations~\cite{Aitchison1991}.

% The oscillating interference pattern in the middle of the WDF clearly explains the intensity profile during the collision.

\section{Conclusion}
In conclusion, we have applied the iterative nonlinear beam propagation approach and the Wigner distribution function to analyze single spatial soliton propagation and the evolution of multi--soliton collisions. The Hamiltonian ray diagrams combined with the phase--space (e.g. Wigner space) description offer a comprehensive and physically intuitive picture of energy evolution in these nonlinear optical phenomena. It is possible to adapt the same approach to other nonlinear optical effects, which is beyond the scope of the current work.

\section*{Acknowledgements}
The authors thank Baile Zhang for useful discussions. Financial support was provided by Singapore's National Research Foundation through the Center for Environmental Sensing and Modeling (CENSAM) and BioSystems and bioMechanics (BioSyM) independent research groups of the Singapore--MIT Alliance for Research and Technology (SMART) Centre, and by the Chevron--MIT University Partnership Program.

%% The Appendices part is started with the command \appendix;
%% appendix sections are then done as normal sections
%% \appendix

%% \section{}
%% \label{}

%% References
%%
%% Following citation commands can be used in the body text:
%% Usage of \cite is as follows:
%%   \cite{key}         ==>>  [#]
%%   \cite[chap. 2]{key} ==>> [#, chap. 2]
%%

%% References with bibTeX database:

\bibliographystyle{elsarticle-num}
\bibliography{SolitonCollisionWDF}

%% Authors are advised to submit their bibtex database files. They are
%% requested to list a bibtex style file in the manuscript if they do
%% not want to use elsarticle-num.bst.

%% References without bibTeX database:

% \begin{thebibliography}{00}

%% \bibitem must have the following form:
%%   \bibitem{key}...
%%

% \bibitem{}

% \end{thebibliography}

\end{document}